\documentclass[]{spie}  %>>> use for US letter paper
\usepackage{amsmath,amsfonts,amssymb}
\usepackage{graphicx}
\usepackage[colorlinks=true, allcolors=blue]{hyperref}
\usepackage{caption}
\usepackage{booktabs}

\title{Fabrication of Pupil Masks for a New Infrared Exoplanet Imager at Keck Observatory}

\author[a,b]{Jialin Li}
\author[c]{Andrew J. Skemer}
\author[c]{Maaike A. M. van Kooten}
\author[d]{Renate Kupke}
\author[d]{Nicholas MacDonald}
\affil[a]{Department of Physics, University of California Santa Cruz, 1156 High St, Santa Cruz, CA 95064, USA}
\affil[b]{Department of Astronomy, University of Arizona, 933 N. Cherry Ave., Tucson, AZ 85718, USA}
\affil[c]{Department of Astronomy and Astrophysics, University of California Santa Cruz, 1156 High St, Santa Cruz, CA 95064, USA}
\affil[d]{University of California Observatory, University of California Santa Cruz, 1156 High Street, Santa Cruz, CA 95112 USA}
\authorinfo{Further author information: %(Send correspondence to A.A.A.)
\\Jialin Li: E-mail: jli428@ucsc.edu
\\Andrew Skemer : E-mail: askemer@ucsc.edu}

\begin{document} 
\maketitle

\begin{abstract}
The Slicer Combined with Array of Lenslets for Exoplanet Spectroscopy (SCALES) is an instrument being designed to perform direct imaging of exoplanets in the mid-infrared (2-5 $\mu$m) with the Adaptive Optics System of W.M. Keck Observatory. To eliminate unwanted thermal infrared radiation, SCALES utilizes both a cold stop for excluding background radiation and a vector vortex coronagraph with Lyot stops for starlight suppression. Optimal geometric masks have been designed. We simulate the propagation of light through the Lyot plane and analyze the on-axis images of stars in the K, L, and M band for the performance of the Lyot stops. Additionally, finalized cold stop and Lyot stop designs are presented along with evaluations on the effects of manufacturing tolerances and tilt in pupil planes caused by off-axis parabolic mirror relays.
\end{abstract}

\keywords{instrumentation, thermal infrared, integral field spectroscopy, adaptive optics}

\section{INTRODUCTION}
\label{sec:intro}
The Slicer Combined with Array of Lenslets for Exoplanet Spectroscopy (SCALES) is an exoplanet imaging spectrograph under construction for the Keck II Adaptive Optics system. SCALES combines the two powerful methods for imaging exoplanets, thermal infrared imaging and coronagraphic integral field spectroscopy that extents to longer wavelengths, covering 2-5 um, to achieve the capability of characterizing cold exoplanet and brown dwarf atmospheres with temperatures less than 600 K.

\begin{figure} [htbp] 
\begin{center}
\includegraphics[height=6cm]{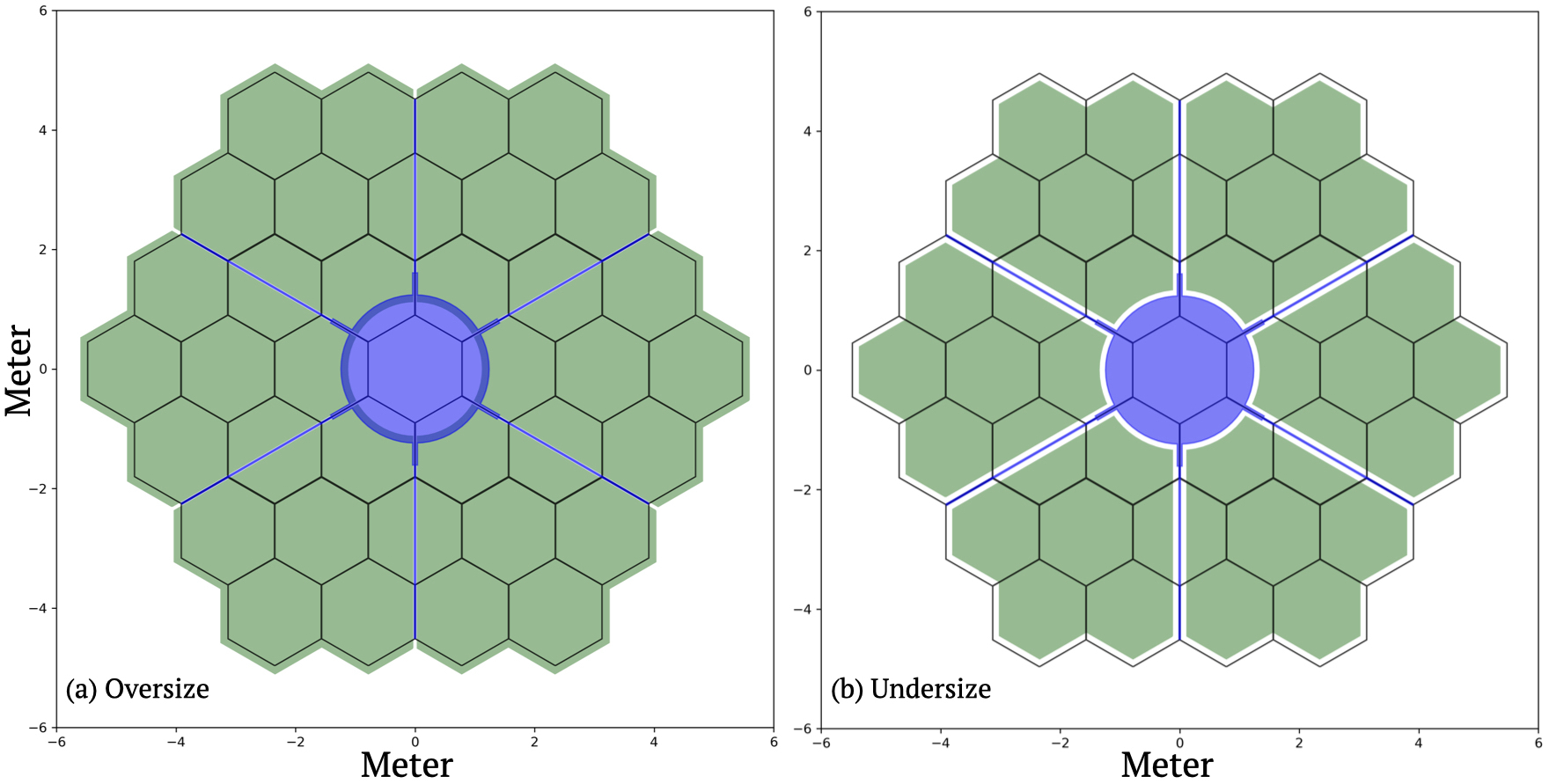}
\end{center}
\caption {The optimal designs of the (a)cold stop and (b)Lyot stop. The secondary structures are shown in blue and the outlines of the primary mirrors are shown in black. The green patches are representations of the carved-out sections in which the light can pass through. Figure adapted from \cite{2021SPIE11823E..1VL}. \label{fig:masks}}
\end{figure}

As SCALES operates in the infrared, cold stops are essential for suppressing unwanted thermal emission from the telescope structure. The cold stop is oversized such that the unwanted radiation from the central obscuration and support structures are blocked out while preserving the maximal amount of signal. A Lyot stop will be placed to stop the diffracted starlight at any wavelengths. The Lyot stop is undersized with oversized inner and supporting structures such that point spread function (PSF) remains stable. Ideally, both masks would match the dimension and shape of the telescope pupil, such that all thermal radiation and scattered light not originated from the primary mirrors can be blocked. However, realistic constraints such as physical fabrication limitations, misalignment, and pupil nutation make the ideal masks unsuitable. Cause by the alignment precision of the Keck adaptive optics K-mirror and telescope beam, the Keck II pupil nutates at approximately 1\% of the diameter of the primary mirror\cite{2016SPIE.9909E..22F}, which leads to misalignment between the pupil and the cold-stop as well as varying point spread function (PSF). One cold stop design and three Lyot stop designs that has taken pupil nutation into account through the modeling of throughput and background emission for SCALES were proposed \cite{2021SPIE11823E..1VL}. The optimal geometric designs for the cold stop and Lyot stop assuming 2\% pupil nutation are shown respectively in the left and right of Figure \ref{fig:masks}.

In this proceeding, we analyzed the ability of starlight suppression of the geometrically optimal Lyot stop designs through simulations of light propagation and finalized the cold and Lyot stop designs. The simulations performed with HCIPy discussed in Section~\ref{sec:hcipy} \cite{2018SPIE10703E..42P}, the effects pupil elongation caused by tilts in optics system is assessed in Section \ref{sec:tiltoptics}, the of the coronagraph slide and Lyot wheel mechanisms and designs are described in \ref{sec:pupilwheel}, the finalized mask designs are presented in \ref{sec:designs}, and our outcomes are summarized in Section \ref{sec:conclusion}. The full SCALES optical design, including performance modeling and analysis, manufacturing and alignment considerations can be found in this proceeding \cite{Kupke2022}. Furthermore, an overview of SCALES and current progress updates can also be found in the same proceeding \cite{Skemer2022}.

\section{HCIPy Simulations and Results} \label{sec:hcipy}
Simulations of diffraction for the Lyot stop designs were carried out through High Contrast Imaging for Python (HCIPy), a Python framework implementing coronagraphy and optical diffraction calculations \cite{2018SPIE10703E..42P}. We analyzed the on-axis images of stars from hexagonal aperture shaped like Keck through a vector vortex chronograph of charge 2 at three wavelengths and generated raw contrast curves. The chosen three wavelengths are the central wavelengths of the K, L, and M bands, which all have a corresponding filter in both the low and medium resolution mode of the IFS. The topological charge of 2 was chosen as that is the  charge of the vortex coronagraph selected for the Keck Planet Imager and Characterizer (KPIC)\cite{2019SPIE11117E..0UP, 2019SPIE11117E..0VE}, which is a series of upgrades for the Keck II adaptive optics system that improves the capabilities of facility instruments like NIRC2. Additionally, higher even values of charge indicate a decreased sensitivity to alignment and is required for smaller inner working angles \cite{2009JOptA..11i4022S}. The effects of atmospheric turbulence or post processing are neglected as we are only examining the Lyot stops’ ability to suppress starlight.

\begin{figure} [htbp] 
\begin{center}
\includegraphics[width=11cm]{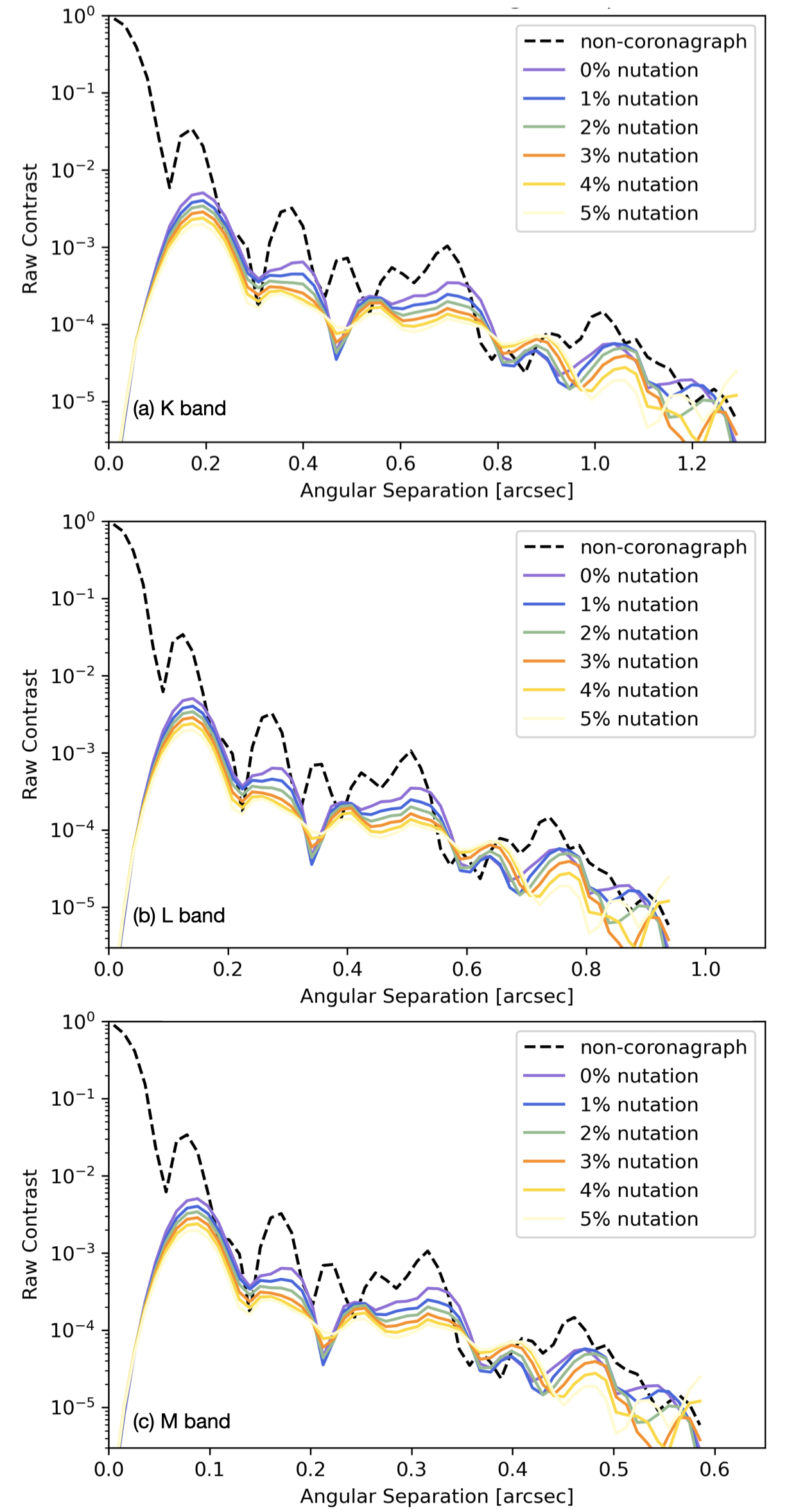}
\end{center}
\caption {Raw contrast vs. angular separation in arcsecs for the (a)K, (b)L, and (c) M band. The no coronagraph case in shown dashed lines and the various undersized coronagraphic masks are shown in solid lines. \label{fig:raw-contrast}}
\end{figure}

A comparison of the raw contrast curves with angular separation in arcsec is shown in Figure \ref{fig:raw-contrast}. The angular separation range increases with wavelength. At most angular separations, the raw contrast with coronagraph is about 10$^1$ higher than that of no coronagraph, suggesting the Lyot Stop designs can effectively suppressing starlight at all bands. Lyot stops with more undersized designs have a sharper contrast at all wavelengths, but it appears to be a marginal effect. The reduction in contrast is likely due to the additional obstruction of light with more conservative designs. Their performance are similar at smaller angular separation and begin diverging from each other at a higher separation.

\section{Fabrication} \label{sec:fabrication}

\subsection{Tilt in Optics}\label{sec:tiltoptics}
\begin{figure} [htbp] 
\begin{center}
\includegraphics[height=6cm]{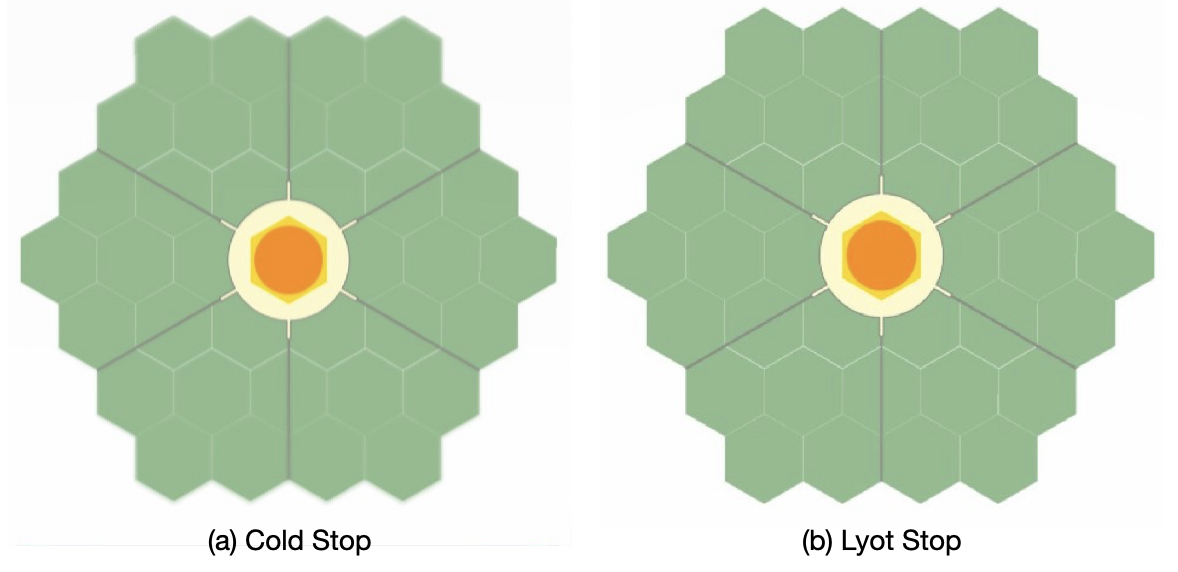}
\end{center}
\caption {Simulated pupil images at the (a) cold stop and (b)Lyot stop of the telescope primary. \label{fig:tiltblur}}
\end{figure}

Due to off-axis parabolic mirrors used to relay the beam throughout the optical path, including in the Keck AO system and within SCALES, the cold stop and Lyot stop pupils are not perfectly symmetric. There is a tilt in the y direction for both cold stop and Lyot stop pupil, with the former reaching 18.75$^\circ$ and latter with 6.84$^\circ$. The simulated pupil images of the telescope primary mirrors at the cold stop and Lyot stop are shown in Figure \ref{fig:tiltblur}. The blurring at the pupil edge for the cold stop is more discernible as the angle of the tilt is higher. Whereas the Lyot stop, the smaller tilt make the edge sharper.

Despite the tilt angle differences, the stretch in the y-direction from four corners of the pupil (left, right, top, and bottom) compared to the central image of the remain the same. The length in the y-axis is measured to be 0.006mm in the central image of the pupil, then it is increased slightly to 0.009mm for images from the left and right corners of the pupil. The change in length is more drastic in the top and bottom images; the images appear to be blurry with sizes 6.785mm at the top and 6.828mm at the bottom. 

To keep the point spread functions symmetric, the cold stop and Lyot stop will still be mounted orthogonal to each other despite the tilt. Undersizing the masks by the size of the blurred images can remove the noise introduced by the blur, but can also lead to loss of throughput. The effect of the tilt, however, is a small effect when compared to pupil nutation caused by the K-mirror rotator. For the case of the cold stop, the undersizing has to be in all directions in order to completely eliminate the blur as it rotates. For 2$\%$ pupil nutation, the optimal cold stop design that takes into account of the tilt would result in the reduction in throughput of 0.125$\%$ when compared to the 2$\%$ oversized cold stop. For the Lyot stop, the design that minimizes the effect of the blur causes a 0.061$\%$ reduction in throughput for all values of nutation. When compare to the impact of the 2$\%$ nutation, which lowers the SNR by about 12$\%$ than the no nutation case, the effect of the pupil tilt is negligible. Thus, we account for the pupil tilt by accepting the slight increased background in having masks sized for the longer direction.

\subsection{Cold Stop Rotating Mechanism and Lyot Wheel Designs}\label{sec:pupilwheel}
The cold stop will be placed in a cold stop rotating mechanism consisting of a stainless steel rotator supported by a aluminium pedestal positioned at the 1st pupil image. With the combination of the the mechanism and a 19 bit resolution encoder, SCALES will be compatible to the field tracking mode of the Keck Adaptive Optics system with accurate position and speed feedback \cite{Banyal2022}. The Lyot stops will reside in the Lyot Wheel with rotary mechanism at the pupil plane, capable of holding up to 15 Lyot stops, neutral density filters, an open position, and a mirror for diverting light into the imaging channel \cite{Skemer2022}. The Lyot wheel utilizes a passive detent mechanism to achieve the tight specifications of pupil mask alignment \cite{2020SPIE11447E..64S}. Details on the instrument architecture and opto-mechanical design can be found in this and previous proceedings \cite{Skemer2022, Banyal2022, Kupke2022, 2020SPIE11447E..64S}.

\subsection{Masks Designs}\label{sec:designs}
\begin{figure} [htbp] 
\begin{center}
\includegraphics[height=\linewidth]{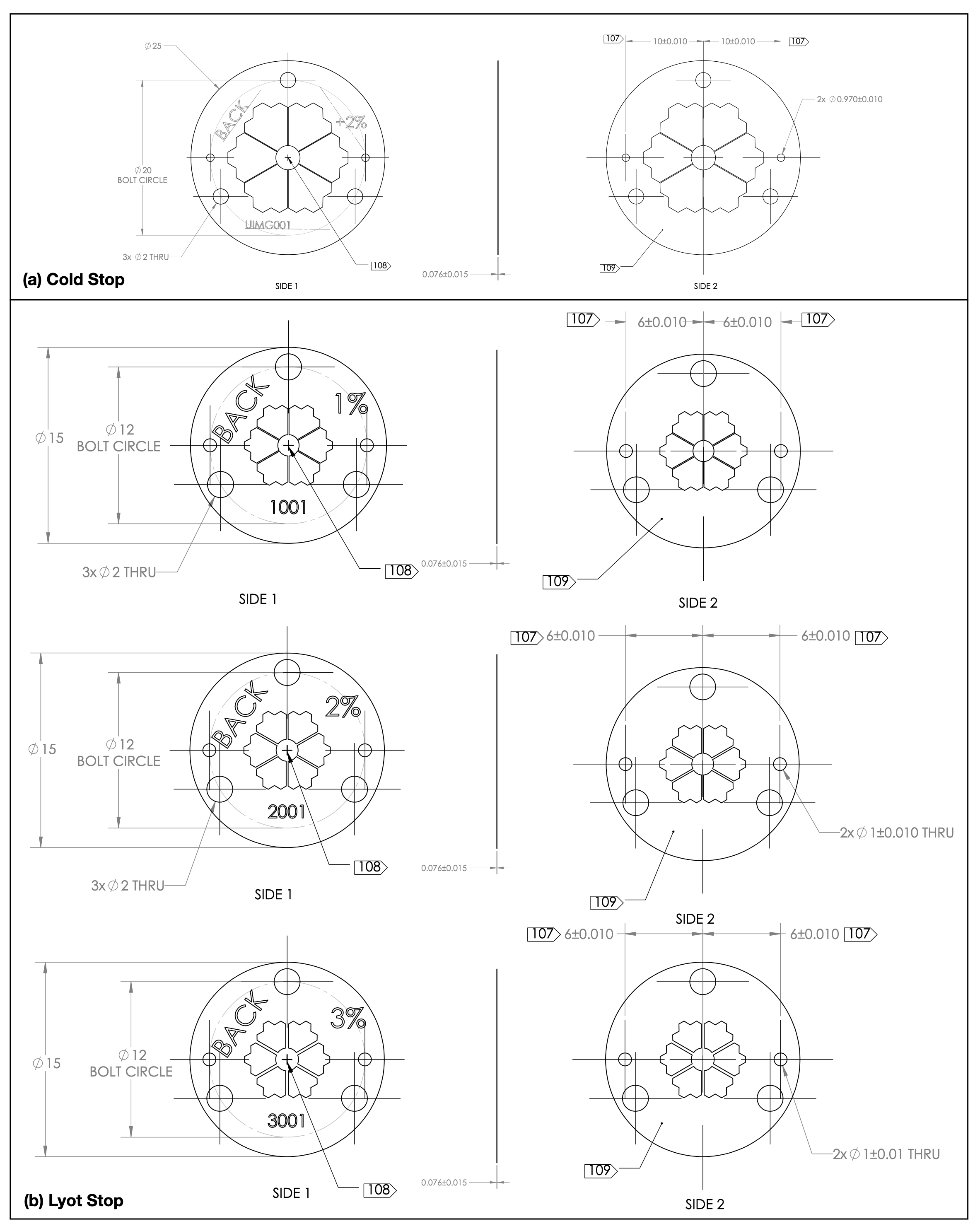}
\end{center}
\caption {CAD designs of the (a)2$\%$ oversized cold stop and (b)1-3$\%$ undersized Lyot stops. \label{fig:cad}}
\end{figure}

Designs of 2$\%$ oversized cold stop and 2$\%$ undersized Lyot stop will be fabricated with modifications due to manufacturing tolerances. A  1$\%$ and 3$\%$ undersized  Lyot stop designs  will be manufactured as well. The former accounts for the most recent measured value of pupil nutation \cite{2016SPIE.9909E..22F}, while the latter is a design with an even more conservative value of nutation. The key dimensions of designs at their respectful pupil plane are listed in Table \ref{tab:dimensions} and CAD designs of each is shown in Figure \ref{fig:cad}. 

The masks will be lithographically etched on stainless steel material and coated with a custom Vanta Black coating. The minimal material thickness is 0.0762 mm for our preferred material, 300 series stainless steel, with a tolerance of $\pm$0.0254 mm during the etching process. This value will be used as the optimal width of spider arm for the cold stop as the design entails the thinnest width. Such value is less a constraint to the Lyot stops as the desired thickness is above this lower limit.

The manufacturer can select the material thickness with the tolerance of $\pm$0.025 depending on their preferred stock size. Once the material thickness can be quoted, the tolerance for the minimal thickness will be well controlled to $\pm$0.005 mm. To ensure the spider widths stays comparable to the expected value, we will produce multiple copies of the same mask design. Furthermore, if the spider arm widths do vary, duplicates of the same masks can be placed as the Lyot wheel can hold up to 15 masks. 

In addition to the actual masks themselves, a few extra features will also be etched as shown in Figure \ref{fig:cad}. Texts will appear on one side of the masks for part identification. A target scribe mark will be placed at the center for alignment of the hexagon aperture to the beam. The three hole pattern, located at the top, bottom left and right corners, is for three \#0-80 fasteners that hold the mask to the holder. Two smaller holes located on the left and right are for pins which locate the mask to its holder. Both sets of holes are outside of the beam.

\begin{table}[]
\centering
\begin{tabular}{cccc}
\hline
Mask            & Outer Radius(mm) & Inner Radius(mm) & Spider Width(mm) \\ \hline
Cold Stop (2\%) & 7.7410        & 1.5565        & 0.0762           \\
Lyot Stop (1\%) & 3.3772        & 0.8145        & 0.0846           \\
Lyot Stop (2\%) & 3.3333        & 0.8459        & 0.1535           \\
Lyot Stop (3\%) & 3.2895        & 0.8835        & 0.2218           \\ \hline
\end{tabular}
\caption{Key dimensions of the masks at their respective pupil planes. The percentage within the parenthesis stands for the pupil nutation value that such mask is designed for.}
\label{tab:dimensions}
\end{table}

Another uncertainty on our designs would be the magnification factor from the telescope primary to the respective pupil planes where the masks are located. We estimated the magnification factor based on the focal length of power optics prior to the pupil planes, each with an expected accuracy of 0.25$\%$. The cold stop and Lyot stop magnification factors are estimated using the root mean square of the focal length tolerances. Three and five powered optics are placed before the cold and Lyot stop, so the magnification factors are 726$\pm$0.40$\%$ and 1596$\pm$0.56$\%$. 

\section{CONCLUSION} \label{sec:conclusion}
We presented in this paper the finalized cold stop and Lyot stop designs for SCALES, a high-contrast lenslet integral field spectrograph and imager operating in the mid-infrared. The finalized designs are the same as the optimal geometric designs with one modification. An arbitrary small value of of 0.08m at the telescope or 0.1102 mm at the pupil was set as the optimal width of the cold stop spider arms to achieve the maximal undersizing of the supporting structures. With the manufacturing tolerances on the minimal thickness being 0.0762$\pm$0.0254 mm at the pupil, which is thinner than the previous arbitrary thickness, this value is used as the optimal width for the cold stop instead. 

Our analysis on the propagation of light simulation through the Lyot plane suggests that our Lyot stop designs are effectively suppressing starlight; the raw contrast of the on-axis star images with a coronagraph in place is increased by a minimum of one order of magnitude at most angular separations and for all three bands we tested. We have determined that the effect of the pupil tilt is small when compared to pupil nutation, thus previous designs will not be altered. We account for the pupil tilt by accepting the slightly increased background in having masks sized for the longer direction. 

\acknowledgments % equivalent to \section*{ACKNOWLEDGMENTS}       
This project was made possible by a Lloyd B. Robinson Undergraduate Research in Instrumentation Award from UC Santa Cruz.  We gratefully acknowledge the support of the Heising-Simons Foundation through grant \#2019-1697. 

% References
\bibliography{main} % bibliography data in report.bib

\begin{thebibliography}{10}

\bibitem{2016SPIE.9909E..22F}
{Femen{\'\i}a Castell{\'a}}, B., {Serabyn}, E., {Mawet}, D., {Absil}, O.,
  {Wizinowich}, P., {Matthews}, K., {Huby}, E., {Bottom}, M., {Campbell}, R.,
  {Chan}, D., {Carlomagno}, B., {Cetre}, S., {Defr{\`e}re}, D., {Delacroix},
  C., {Gomez Gonzalez}, C., {Jolivet}, A., {Karlsson}, M., {Lanclos}, K.,
  {Lilley}, S., {Milner}, S., {Ngo}, H., {Reggiani}, M., {Simmons}, J., {Tran},
  H., {Vargas Catalan}, E., and {Wertz}, O., ``{Commissioning and first light
  results of an L'-band vortex coronagraph with the Keck II adaptive optics
  NIRC2 science instrument},'' in [{\em Adaptive Optics Systems
  V}{\nolinebreak\hspace{0.1em}]},  {Marchetti}, E., {Close}, L.~M., and
  {V{\'e}ran}, J.-P., eds., {\em Society of Photo-Optical Instrumentation
  Engineers (SPIE) Conference Series} {\bf 9909},  990922 (July 2016).

\bibitem{2021SPIE11823E..1VL}
{Li}, J. and {Skemer}, A., ``{Cold-stop and lyot stop designs for a new
  Infrared Exoplanet Imager at Keck Observatory},'' in [{\em Society of
  Photo-Optical Instrumentation Engineers (SPIE) Conference
  Series}{\nolinebreak\hspace{0.1em}]},  {\em Society of Photo-Optical
  Instrumentation Engineers (SPIE) Conference Series} {\bf 11823},  118231V
  (Sept. 2021).

\bibitem{2018SPIE10703E..42P}
{Por}, E.~H., {Haffert}, S.~Y., {Radhakrishnan}, V.~M., {Doelman}, D.~S., {van
  Kooten}, M., and {Bos}, S.~P., ``{High Contrast Imaging for Python (HCIPy):
  an open-source adaptive optics and coronagraph simulator},'' in [{\em
  Adaptive Optics Systems VI}{\nolinebreak\hspace{0.1em}]},  {Close}, L.~M.,
  {Schreiber}, L., and {Schmidt}, D., eds., {\em Society of Photo-Optical
  Instrumentation Engineers (SPIE) Conference Series} {\bf 10703},  1070342
  (July 2018).

\bibitem{Kupke2022}
Kupke, R., Stelter, R.~D., Hasan, A., Surya, A., Kain, I., Briesemeister, Z.,
  Li, J., Hinz, P., Skemer, A., Gerard, B., Dillon, D., and Ratliff, C.,
  ``Scales for keck: Optical design,'' in [{\em Ground-based and Airborne
  Instrumentation for Astronomy IX}{\nolinebreak\hspace{0.1em}]},   {\bf
  12184}, SPIE (2022).

\bibitem{Skemer2022}
{Skemer}, A.~J., {Stelter}, R.~D., {Sallum}, S., {MacDonald}, N., {Kupke}, R.,
  {Ratliff}, C., {Banyal}, R.~K., {Sivarani}, T., {Fitzgerald}, M.~P.,
  {Kassis}, M., {Absil}, O., {Alvarez}, C.~A., {Batalha}, N., {Boucher}, M.-A.,
  {Bourgenot}, C., {Briesemeister}, Z., {Deich}, W., {Divakar}, D., {Filion},
  G., {Gauvin}, {\'E}., {Gonzales}, M., {Greene}, T., {Hasan}, A., {Hinz}, P.,
  {Jensen-Clem}, R., {Johnson}, C., {K. V.}, Govinda, K.~I., {Kruglikov}, G.,
  {Lach}, M., {Landry}, J.-T., {Li}, J., {Lyke}, J., {Magone}, K., {Marin}, E.,
  {Martin}, E., {Martinez}, R., {Mawet}, D., {Miles}, B., {Parkash}, A.,
  {Sandford}, D., {Sethuram}, R., {Sheehan}, P., {Sohn}, J., {Surya}, A.,
  {Varshney}, H., and {Wang}, E.

\bibitem{2019SPIE11117E..0UP}
{Pezzato}, J., {Jovanovic}, N., {Mawet}, D., {Ruane}, G., {Wang}, J.,
  {Wallace}, J.~K., {Colborn}, J.~K., {Cetre}, S., {Bond}, C.~Z., {Bartos}, R.,
  {Calvin}, B., {Delorme}, J.-R., {Echeverri}, D., {Jensen-Clem}, R., {McEwen},
  E., {Lilley}, S., {Wetherell}, E., and {Wizinowich}, P., ``{Status of the
  Keck Planet Imager and Characterizer phase II development},'' in [{\em
  Society of Photo-Optical Instrumentation Engineers (SPIE) Conference
  Series}{\nolinebreak\hspace{0.1em}]},  {\em Society of Photo-Optical
  Instrumentation Engineers (SPIE) Conference Series} {\bf 11117},  111170U
  (Sept. 2019).

\bibitem{2019SPIE11117E..0VE}
{Echeverri}, D., {Ruane}, G., {Jovanovic}, N., {Hayama}, T., {Delorme}, J.-R.,
  {Pezzato}, J., {Bond}, C., {Wang}, J., {Mawet}, D., {Wallace}, J.~K., and
  {Serabyn}, E., ``{The vortex fiber nulling mode of the Keck Planet Imager and
  Characterizer (KPIC)},'' in [{\em Society of Photo-Optical Instrumentation
  Engineers (SPIE) Conference Series}{\nolinebreak\hspace{0.1em}]},  {\em
  Society of Photo-Optical Instrumentation Engineers (SPIE) Conference Series}
  {\bf 11117},  111170V (Sept. 2019).

\bibitem{2009JOptA..11i4022S}
{Swartzlander}, Grover~A., J., ``{The optical vortex coronagraph},'' {\em
  Journal of Optics A: Pure and Applied Optics}~{\bf 11},  094022 (Sept. 2009).

\bibitem{Banyal2022}
Banyal, R.~K., Hasan, A., Kupke, R., Sivarani, T., Skemer, A.~J., MacDonald,
  N., Sallum, S., Deich, W., Divakar, D.~K., Fitzgerald, M., Govinda, K.~V.,
  Prakaesh, A., Ratliff, C., Sethuram, R., Sriram, S., Stelter, D., Surya, A.,
  Varshney, H.~M., and Wang, E., ``Design of an ir imaging channel for the keck
  observatory scales instrument,'' in [{\em Advances in Optical and Mechanical
  Technologies for Telescopes and Instrumentation
  V}{\nolinebreak\hspace{0.1em}]},   {\bf 12188}, SPIE (2022).

\bibitem{2020SPIE11447E..64S}
{Stelter}, R.~D., {Skemer}, A.~J., {Sallum}, S., {Kupke}, R., {Hinz}, P.,
  {Mawet}, D., {Jensen-Clem}, R., {Ratliffe}, C., {MacDonald}, N., {Deich}, W.,
  {Kruglikov}, G., {Kassis}, M., {Lyke}, J., {Briesemeister}, Z., {Miles}, B.,
  {Gerard}, B., {Fitzgerald}, M., {Brandt}, T., and {Marois}, C., ``{Update on
  the preliminary design of SCALES: the Santa Cruz Array of Lenslets for
  Exoplanet Spectroscopy},'' in [{\em Society of Photo-Optical Instrumentation
  Engineers (SPIE) Conference Series}{\nolinebreak\hspace{0.1em}]},  {\em
  Society of Photo-Optical Instrumentation Engineers (SPIE) Conference Series}
  {\bf 11447},  1144764 (Dec. 2020).

\end{thebibliography}
\bibliographystyle{spiebib} % makes bibtex use spiebib.bst

\end{document}